\documentstyle[preprint,aps,epsfig]{revtex}
\tighten
\def\bea{\begin{eqnarray}}
\def\beann{\begin{eqnarray*}}
\def\beq{\begin{equation}}
\def\eea{\end{eqnarray}}
\def\eeann{\end{eqnarray*}}
\def\eeq{\end{equation}}
\def\nn{\nonumber}

\newcommand{\br}{\bbox{r}}

\newcommand{\bcdot}{\bbox{\cdot}}

\newcommand{\bq}{\bbox{q}}

\newcommand{\bL}{\bbox{L}}

\newcommand{\bS}{\bbox{S}}

\newcommand{\bsigma}{\bbox{\sigma}}
\newcommand{\bnabla}{\bbox{\nabla}}

\newcommand{\btau}{\bbox{\tau}}

\begin{document}
\draft
\title{Long- and medium-range components of the nuclear force in quark-model 
based calculations}
\author{
D. Hadjimichef$^{1}$, J. Haidenbauer$^{2}$, and G. Krein$^{3}$ \\
{\small $^1$ Departamento de F\'{\i}sica, Universidade Federal de Pelotas, 
96010-900 Pelotas, RS, Brazil} \\
{\small $^2$ Forschungszentrum J\"ulich, Institut f\"ur Kernphysik, 
D-52425 J\"ulich, Germany} \\
{\small $^3$ Instituto de F\'{\i}sica Te\'{o}rica, Universidade Estadual 
Paulista } \\
{\small Rua Pamplona, 145 - 01405-900 S\~{a}o Paulo, SP, Brazil} 
}
\maketitle
\begin{abstract}
Quark-model descriptions of the nucleon-nucleon interaction contain two
main ingredients, a quark-exchange mechanism for the short-range repulsion 
and meson-exchanges for the medium- and long-range parts of the interaction. 
We point out the special role played by higher partial waves, and in 
particular the $^1F_3$, as a very sensitive probe for the meson-exchange 
part employed in these interaction models. 
In particular, we show that the presently available models fail to 
provide a reasonable description of higher partial waves and 
indicate the reasons for this shortcoming. 
\end{abstract}
\vspace{2.5cm}
\noindent{PACS NUMBERS: 21.30.-x, 13.75.Cs, 24.85.+p, 12.39.Jh}

\vspace{1.0cm}

\newpage 
\section{Introduction}

The traditional, and most accurate description of the nucleon-nucleon ($NN$) 
force at low energies is based on meson-exchange models. There are many 
versions of such models in the literature (cf., e.g., Refs.~\cite{Bonn,Bonn2} 
for a short historical view and many references).  With almost no 
exception, the various models have the common feature that the long-range 
part of the interaction is described by one-pion exchange (OPE) and the
medium-range part is described by contributions from two-pion exchange, 
usually parametrized in terms of the $\rho$ and $\sigma$ mesons.   
On the other hand, the treatment of the short-range part of the 
interaction can differ considerably among the $NN$ models. 
This part is assumed to receive contributions from multi-meson exchanges. 
At very short distances the interaction is either parametrized 
phenomenologically or regularized by means of vertex form factors. 
Those parametrizations or the form factors are expected to be explained 
ultimately by invoking quark-gluon degrees of freedom. 

However, the direct use of the QCD Lagrangian (or Hamiltonian) for studying
processes at the nuclear scale has been so far possible only in large-scale 
numerical simulations on a supercomputer. The use of a quark model
seems therefore necessary for analytical calculations. Unfortunately, the 
formulation of an accurate, and at the same time sufficiently simple quark 
model is very difficult, for several reasons. Perhaps the most notorious 
obstacle 
is our difficulty in identifying the relevant effective degrees of freedom 
that operate at the confinement scale. Despite of this, a large body of 
hadronic spectroscopic and strong-decay data can be described reasonably well
by the constituent quark model (CQM)~\cite{LeYaouanc}. In the CQM, the low 
energy spectrum of QCD is postulated to be built from spin-1/2 colored 
{\em constituent} massive quarks, which are confined within hadrons and 
interact weakly through one-gluon exchange (OGE). 
 
Motivated by its simplicity and relative success in describing the 
data, many authors have used the CQM to study the short-range part of the $NN$ 
interaction in terms of the OGE, using different approaches for the motion of 
the six-quark system. In such schemes, the $NN$ repulsion at short 
distances is 
generated dominantly by the quark Pauli exclusion principle and the color 
hyperfine interaction of the OGE. The initial works were based on adiabatic
approximations of the Born-Oppenheimer type. The work of Liberman is the first
on these lines~\cite{Libermann}, followed by the ones by Neudatchin et 
al.~\cite{Neudatchin} and Harvey~\cite{Harvey}. Beyond the adiabatic 
approximation, the resonating group method (RGM) has been widely used. 
Here the pioneering works stem from Ribeiro~\cite{Ribeiro}, Warke and 
Shankar~\cite{WarkeShanker}, Oka and Yazaki~\cite{OkaYazaki}, and Faessler
et al.~\cite{Faessler}. 

A common characteristics of these calculations is that they are unable to 
describe the qualitative features of the long- and medium-range 
parts of the $NN$ interaction. 
In particular, they fail to describe the observed 
spin-orbit splitting of the spin-triplet $P$-wave phase shifts. 
In order to accommodate these features,
meson exchanges and/or phenomenological potentials are added 
to the OGE. First of all the OPE interaction 
is taken into account. In addition, some medium-ranged attractive 
contributions are supplemented. For example, in the works of the 
T\"ubingen-Salamanca (TUEB-SAL)~\cite{TUEB-SAL,TUEB1} and the 
Salamanca-Valencia (SAL-VAL)~\cite{SAL-VAL} groups the exchange
of a $\sigma$ meson is introduced.
The model developed by the Tokyo group (TOK)~\cite{TOK} contains,
besides $\pi$ and $\sigma$ exchange, an additional attractive
phenomenological potential with different strength for each
spin-isospin channel. 
In the model of the Kyoto-Niigata group (KYO-NII)~\cite{KYO-NII}, 
in addition to $\pi$ and $\sigma$, 
all other members of the scalar and pseudoscalar SU(3) meson nonets are 
included in an attempt to describe simultaneously nucleon-nucleon and 
hyperon-nucleon data. 
A common characteristics of these models is that 
vector-meson exchanges ($\omega$, $\rho$) are not considered. The 
rational for this being that the interactions generated by $\omega$ 
and $\rho$ exchange are presumed to be of very short range and therefore 
their effects should be more 
appropriately taken into account by a quark-exchange mechanism. Another reason
for leaving the vector mesons out is that vector meson exchange between 
quarks of different nucleons provide contributions qualitatively similar to 
the ones provided by the quark-exchange mechanism and the simultaneous 
consideration of both contributions would therefore lead to double 
counting~\cite{Yaz}. 

In all these approaches the additional parameters, such as meson-quark
coupling strengths and form factors, are adjusted in part 
by a fit to the low $NN$ partial waves, i.e. those partial waves that 
are mostly sensitive to the short-range part of the $NN$ interaction. 
In general, the 
resulting description of the $NN$ phase shifts, in particular of the 
$S-$ and $P-$waves, is very impressive. This is certainly an achievement
because it is important to realize that the calculations are heavily 
constrained by the requirement that the added interactions still give a 
decent description of the mass splittings of the low-lying baryonic spectrum.
This remark is particularly relevant for those approaches where the 
meson-exchange pieces contribute also to isolated 
baryons \cite{TUEB-SAL,TUEB1,SAL-VAL,KYO-NII} and not only to the
$NN$ interaction \cite{TOK}. 

Higher partial waves are predominantly 
determined by the longer ranged pieces of the $NN$ force. 
These partial waves are usually not considered in the fitting
procedure and therefore the corresponding results can be
regarded as genuine predictions. In particular, this 
means that those higher partial waves are a good testing ground 
for the reliability of the medium- and long-range components 
employed in those quark models of the $NN$ interaction. 
In practice, however, the predictions of quark models for higher 
partial waves are rarely shown. 
There are only a few works where the authors present 
phase shifts for $F$- \cite{KYO-NII} or even $G$- waves \cite{TOK}. 
Indeed the results are not very encouraging! 
They reveal striking differences from the phase-shift
analyses but also from the phases predicted by conventional 
meson-exchange models of the $NN$ interaction. 

In the present paper we want to investigate the origin of these differences. 
Specifically we want to examine the ingredients that constitute
the medium- and long-range pieces of quark models and compare them
with those used in conventional meson-exchange models. 
Thereby we aim at a qualitative appreciation of the reasons for the 
observed failure in describing the higher $NN$ partial waves 
in terms of the dynamics on which those quark models are based. 
Thus, our study is complementary to a 
recent investigation carried out by the Paris group~\cite{Paris-QM}. 
In this work $NN$ observables were calculated with a model 
built from the core (short-range) part of the quark model of the 
Tokyo group \cite{TOK} and supplemented, at intermediate
and long internucleonic distances, by the $NN$ forces generated 
from the Paris potential~\cite{Paris-NN}.
It was found that such an approach leads to a very poor description 
of the data, with $\chi^2/$data ranging from 20 to 160. 

Our paper is organized as follows. In the next section we review
shortly the ingredients of those quark models of the $NN$ interaction
that we consider in our investigation. Furthermore we argue and
establish via sample calculations that the $F$ and $G$ waves are not 
sensitive anymore to the short-range part of the $NN$ force, governed
by quark-exchange mechanisms, and therefore are very well
suited for testing the medium-range pieces that are employed in 
present-days quark models. 
In Sect. III we compare the predictions of specific quark models 
for the $^1F_3$ and $^1G_4$ partial waves with those of a simple
conventional one-boson exchange model of the $NN$ force. In
addition we carry out a detailed analysis of the behavior
of the corresponding potentials for internucleonic distances
around 1 fm in order to understand the dynamical origin of
the differences that we observe in the phase-shift results. 
The paper ends with a general discussion about possible origins for
the failure of quark models in describing those higher partial waves.
Furthermore suggestions on a different strategy to study the 
short-range part of the $NN$ force as derived from subnucleonic
degrees of freedom are given.

\section{Quark exchange and higher $NN$ partial waves}

The medium-range parts of models for the $NN$ forces can be 
investigated most efficiently by looking at higher partial waves of 
the $NN$ interaction \cite{Bonn}. For orbital angular
momenta $L \geq 3$ ($F$, $G$, etc. waves) the centrifugal barrier
is, in general, already sufficiently large to suppress contributions
from the short-range part of the $NN$ interaction, specifically from
quark-exchange processes, as we will show below. Furthermore, it
is preferable to look at spin-singlet partial waves because here the
strong tensor force from the OPE is absent and possible
spin-orbit forces cannot contribute either. 
These contributions to the $NN$ interaction are not relevant for the 
points we want to address. From those considerations
it follows that the $^1F_3$ should be the best candidate
for testing models for the medium-range interaction and most of
our study will concentrate on this partial wave. However, we will
look at the $^1G_4$ as well. 

Our aim in this section is to demonstrate explicitly that the $F$ waves 
are indeed
relatively insensitive to the short-ranged pieces of the $NN$ interaction,
i.e. those that involve quark exchanges between the nucleons. 
For that purpose we solve the scattering equation (Schr\"odinger equation) 
for some quark models using, however, {\it only}
the part of the effective $NN$ interaction without the pieces that involve
quark exchange, and compare the resulting phase shifts with those
obtained for the complete $NN$ interaction model that include quark
exchange. Specifically, we solve 
\beq
\Biggl[-\frac{\bnabla^2}{M} + V^{D}_{NN}(\br)\Biggr]\psi(\br) = E \psi(\br) ,
\label{effSchr}
\eeq
where $M$ is the nucleon mass, $E$ is the two-nucleon relative energy, and
$V^D_{NN}$ is the ``direct'' effective $NN$ interaction kernel. The 
``exchange'' contribution to the effective $NN$ interaction is neglected.

In the case of the TOK potential, $V^D_{NN}(\br)$ is the effective meson
exchange potential (EMEP) $\bar V^{\text{EMEP}}$ whose explicit form is 
given by Eqs.~(16) through (26) in Ref.~\cite{TOK}. 
It contains contributions from the OPE, from a $\sigma$-like part and 
from an attractive phenomenogical central, spin- 
and isospin-dependent, potential of Gaussian form. Note that the 
$\pi$ and $\sigma$ exchanges take place between the quarks. 
The corresponding contribution to $V^{D}_{NN}(\br)$ is the Fourier
transform of the convolution of the microscopic quark-quark interaction 
$V_{qq}(\bq)$ and the nucleon form factor $F(\bq)$ at each vertex
\beq
V^{D}_{NN}(\br) = \int \frac{d\bq}{(2\pi)^3}\,e^{i\bq\bcdot\br}\, 
F(\bq)\,V_{qq}(\bq)\,F(\bq) ,
\label{VD}
\eeq
cf. their Eq. (19). We also want to mention that their 
pion-exchange contribution contains a quadratic spin-orbit term 
of the form $-\bar V_{QSO}\,\bsigma\bcdot\bsigma\,\bL^2$, 
cf. Eqs.~(18) and (25) of \cite{TOK}, which does not vanish
for singlet states. Our calculations are based on the model $Q$
as specified in Table 2 of Ref.~\cite{TOK}. 

The KYO-NII potential contains $\pi$ exchange as well as the 
exchange of two scalar (SU(3) flavor-singlet and octet) mesons. 
All mesons are exchanged between the quarks. The quark-quark interactions,
$V_{qq}$, are simply the standard one-boson-exchange potentials for 
the $\pi$ and scalar mesons, respectively. 
The effective meson exchange potential $V^{D}_{NN}(\br)$ is obtained
via a convolution according to Eq.~(\ref{VD}). 
In our calculation we employ the model $FSS$
as specified in Table III of Ref.~\cite{KYO-NII}. 
 
The TUEB-SAL potential includes the $\pi$ and $\sigma$ mesons; 
both are exchanged between the quarks. The explicit form of
their quark-quark interactions can be found, e.g., in Ref.~\cite{TUEB1}.
The effective meson exchange potential $V^{D}_{NN}(\br)$ is again 
obtained via a convolution according to Eq.~\ref{VD}. 
Our calculation are based on the model parameters that were 
employed in Ref.~\cite{TUEB1}. 
 
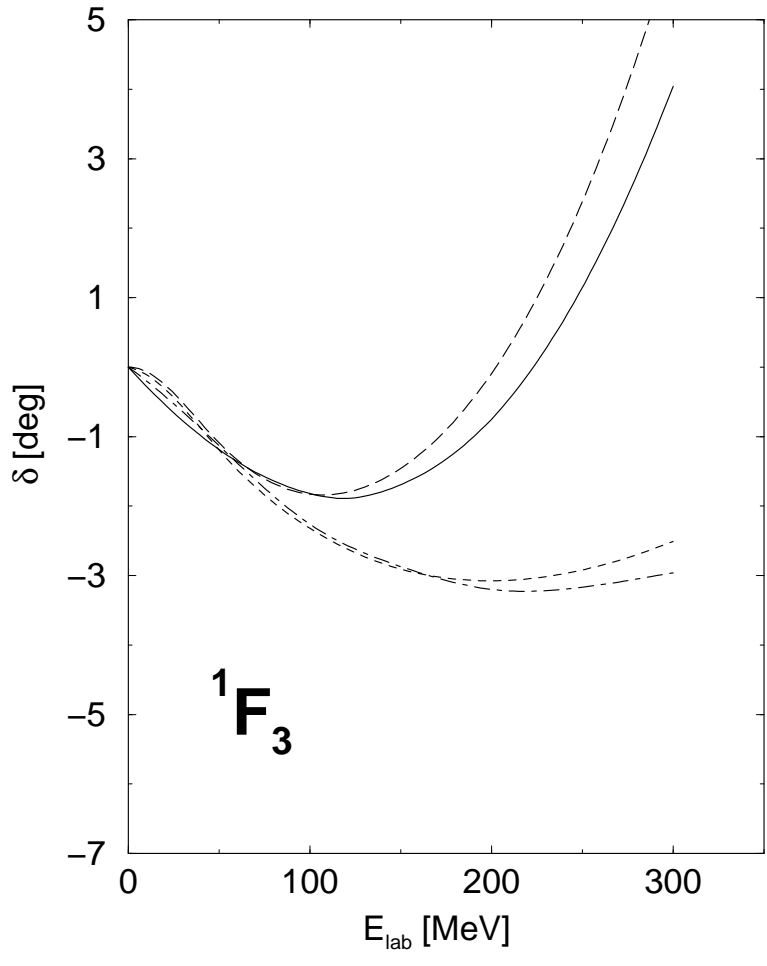
\begin{figure}
\caption{$^1F_3$ phase shift. Comparison of the results of the 
Tokyo \protect\cite{TOK} (solid line) and Kyoto \protect\cite{KYO-NII} 
(dash-dotted line) groups based on the full model with 
our calculation (dashed curves) in which only the ``direct'' part of
the effective $NN$ interaction is employed, cf. Sect. II.}
\label{phcom} 
\end{figure}

Results for the $^1F_3$ phases are presented in Fig.~\ref{phcom}.
The solid and dash-dotted lines show the phase shifts of the 
complete calculation with the TOK and KYO-NII potentials, 
respectively, taken from the original works \cite{TOK,KYO-NII}.
The dashed curves are corresponding results obtained by us. 
As mentioned above, in our calculation only the medium- and 
long-range part of these potentials was taken into account. 
Short-range contributions from the quark-exchange processes were 
omitted. Evidently, the differences between the two calculations
are fairly small, which means that the $^1F_3$ phase shift is indeed 
primarily determined by the medium- and long-range part of the 
$NN$ interaction. The quark-exchange part has definitely still an 
influence on this phase, but only in a quantitative sense and not
on its qualitative behavior. 

Note that we have carried out similar calculations also for other 
quark models of the $NN$ interaction such as the TUEB-SAL and SAL-VAL 
potentials. Specifically, for the TUEB-SAL model phase-shift results 
were provided privately to us by one of the authors of Ref.~\cite{TUEB1} 
and we could check explicitly that also in this case our results agree 
well with theirs. 
 
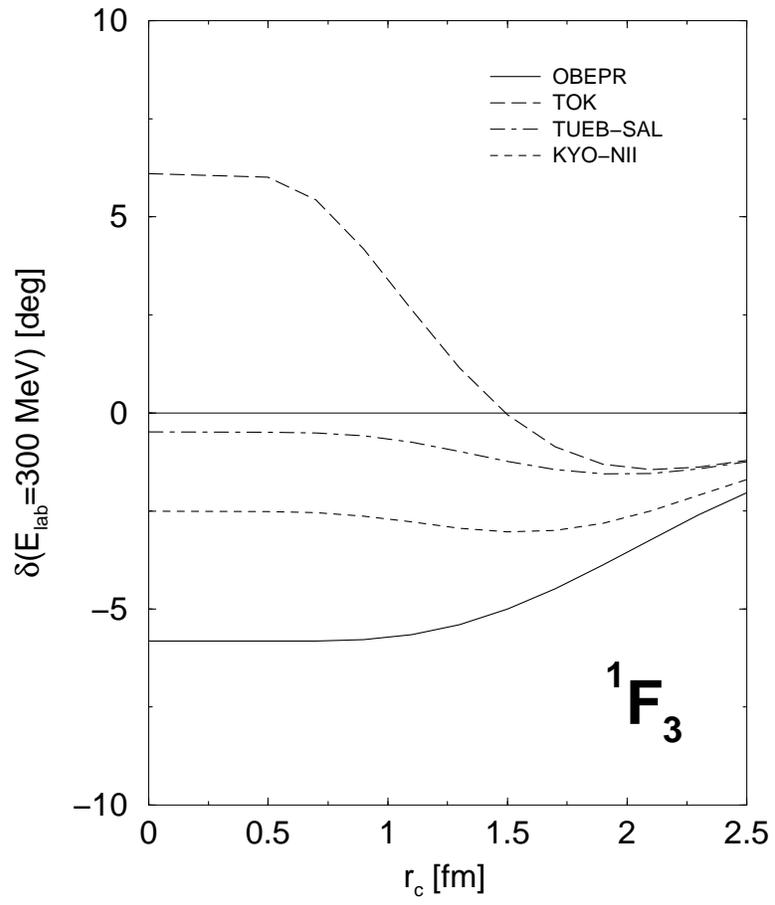
\begin{figure}
\caption{$^1F_3$ phase shifts at $E_{lab}=300$~MeV as a function of 
the cutoff radius $r_{c}$ for the one-boson-exchange
model OBEPR \protect\cite{Bonn} (solid line) and the 
quark models of the Tokyo \protect\cite{TOK} (long-dashed line),
T\"ubingen-Salamanca \protect\cite{TUEB1} (dash-dotted line), and
Kyoto-Niigata \protect\cite{KYO-NII} (short-dashed line) groups.}
\label{phcut} 
\end{figure}

In order to substantiate our conjecture that the $F$ waves are rather 
insensitive to the short-range part of the $NN$ interaction we designed 
a further test. We apply a cutoff of the form
\beq
f(r) = \frac{1}{\left[1 + (r_c/r)^{10}\right]} . 
\label{cutoff}
\eeq
to the $NN$ potential $V^{D}_{NN}$.
This cutoff function acts like a step function, such that for distances
$r$ smaller than $r_c$, $f(r)$, and therefore the $NN$ potential, is
practically zero. Then we insert this modified potential into the
Schr\"odinger equation, calculate the phase shifts at a fixed
energy and study their dependence on the cutoff radius $r_c$. 
Corresponding results for the $^1F_3$ partial wave at $E_{lab}=300$~MeV,
based on several $NN$ interaction models, are shown in Fig.~\ref{phcut} 
as a function of the cutoff radius $r_c$. One sees that the results for 
this partial wave are, in general, rather insensitive to the cutoff 
radius - and accordingly to the $NN$ interaction - 
for values of $r_c$ smaller than $r_c\approx~1$~fm. Only in case of 
the TOK potential there is a somewhat larger sensitivity resulting 
in deviations of the order of $10\%$ already for $r_c\approx~0.7$~fm.

Similar features were found also for the $^3F_3$ partial wave. For
$G$ waves (and in particular the $^1G_4$) it turned out that the 
phase shifts are even insensitive to the $NN$ interaction for 
internucleon distances up to $r_c\approx~1.5$~fm. 

Let us come back to Fig.~\ref{phcut} again. With increasing cutoff
radius $r_c$ much of the medium-range contributions will be suppressed
as well and only the long-range part will be left, which is in case 
of the $^1F_3$ the spin-spin part of the OPE. 
Its contribution is present in all considered $NN$ potentials and 
therefore the phase-shift results should all converge to a common value 
for increasing values of $r_c$. However, even at the highest
value shown in Fig.~\ref{phcut}, $r_c$ = 2.5 fm, there are still
descrepancies. They are partly due to differences in 
the pion coupling constant and regularization schemes employed in
the considered $NN$ models. But primarily they indicate that the
medium-range part of those $NN$ interaction models is still 
sizable, even at internucleonic distances $r \approx$ 2.5 fm. 

\section{Medium-range meson-exchange and higher $NN$ partial waves}

Having established the insensitivity of $F$- and higher partial waves
to the quark-exchange part of the effective $NN$ interaction, we examine 
in this section the performance of the different quark models in 
describing these phase shifts. Furthermore we scrutinize the 
dynamical ingredients that constitute the medium-range part of those 
interaction models. Specifically, we analyze the features of 
these potentials in r-space and we compare them
with conventional meson-exchange models of the $NN$ interaction.
For the latter we take the r-space version (OBEPR) of the Bonn $NN$ 
model~\cite{Bonn}. There are certainly much more refined $NN$ models
in the literature - in terms of the dynamical input (e.g, the full Bonn
model~\cite{Bonn}) as well as with regard to the description of $NN$
phase shifts \cite{nijm,cdbonn}. However, for the qualitative 
comparison that we have in mind we need a model that has practically
no non-localities and therefore is easy to handle in r-space. 
Furthermore, the Bonn OBEPR model includes all the one-boson-exchange
contributions ($\pi$, $\rho$, $\omega$, $\sigma$, ..., exchanges) 
that are usually present in meson-exchange models and, most
importantly, yields a fair description of the higher partial waves
that we want to study. Therefore, the model OBEPR is indeed very 
well suited for our purpose. 

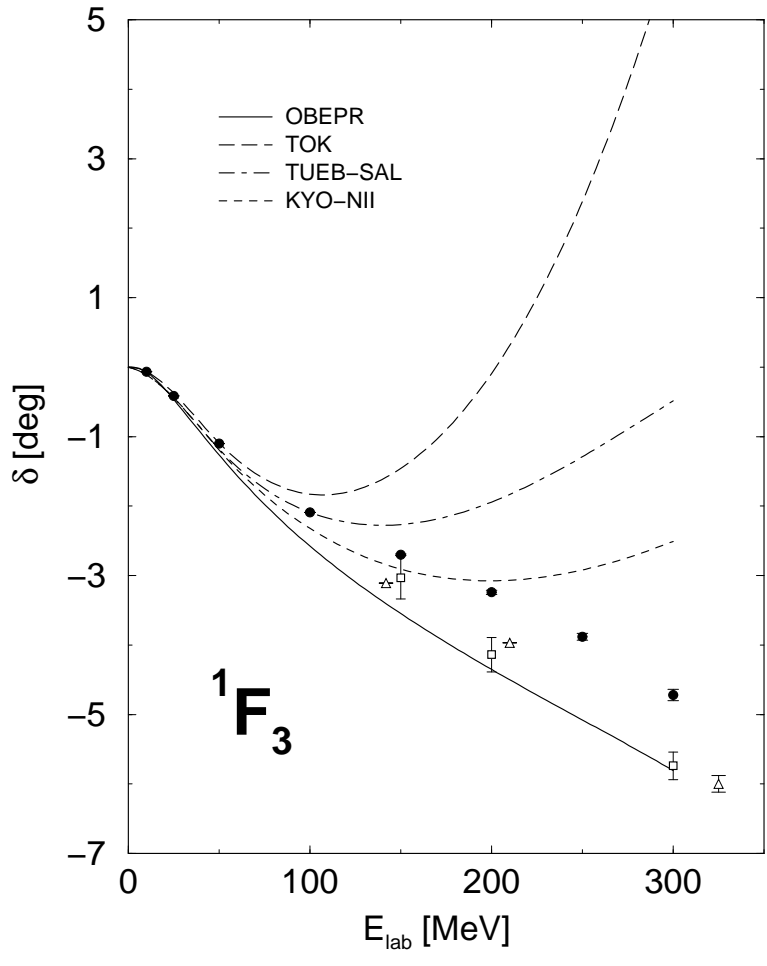
\begin{figure}
\caption{$^1F_3$ phase shifts predicted by the considered potential 
models. Same description of curves as in Fig.~\ref{phcut}. 
Experimental phase shifts are from the analyses of the 
Nijmegen group \protect\cite{exper1} (solid circles),
Arndt et al. \protect\cite{exper2} (squares), and 
Bugg et al. \protect\cite{exper3} (triangles).} 
\label{ph1f3} 
\end{figure}

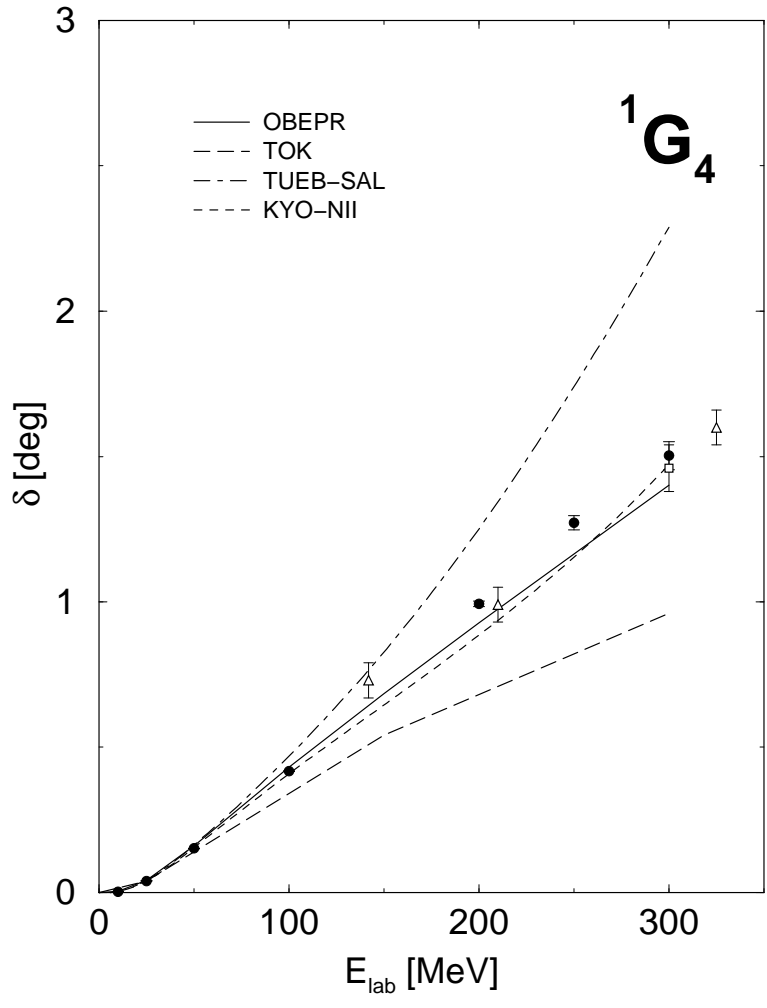
\begin{figure}
\caption{$^1G_4$ phase shifts. Same description as in 
Fig.~\ref{ph1f3}.}
\label{ph1g4} 
\end{figure}

In Figs.~\ref{ph1f3} and \ref{ph1g4} we show results for the 
$^1F_3$ and $^1G_4$ waves, respectively, as a function 
of the $NN$ laboratory energy. The data points are taken from 
the phase shift analyses of Refs.~\cite{exper1,exper2,exper3}. 
Evidently, the $^1F_3$ phase shifts predicted by
the quark models differ significantly from the one of the 
conventional meson-exchange model OBEPR, cf. Fig.~\ref{ph1f3}. 
Specifically, the latter provides a reasonable description
of this partial wave whereas the quark models deviate 
strongly from the experimental results. In fact, the KYO-NII 
potential is at least still in qualitative agreement with the data 
whereas the TOK potential yields completely unrealistic 
results. The predictions of the latter even change sign at higher 
energies. Indeed all quark-model results show an upwards-rising of 
the $^1F_3$ phase shift for higher energies. This indicates
that the medium-range part of all these models is too attractive.
 
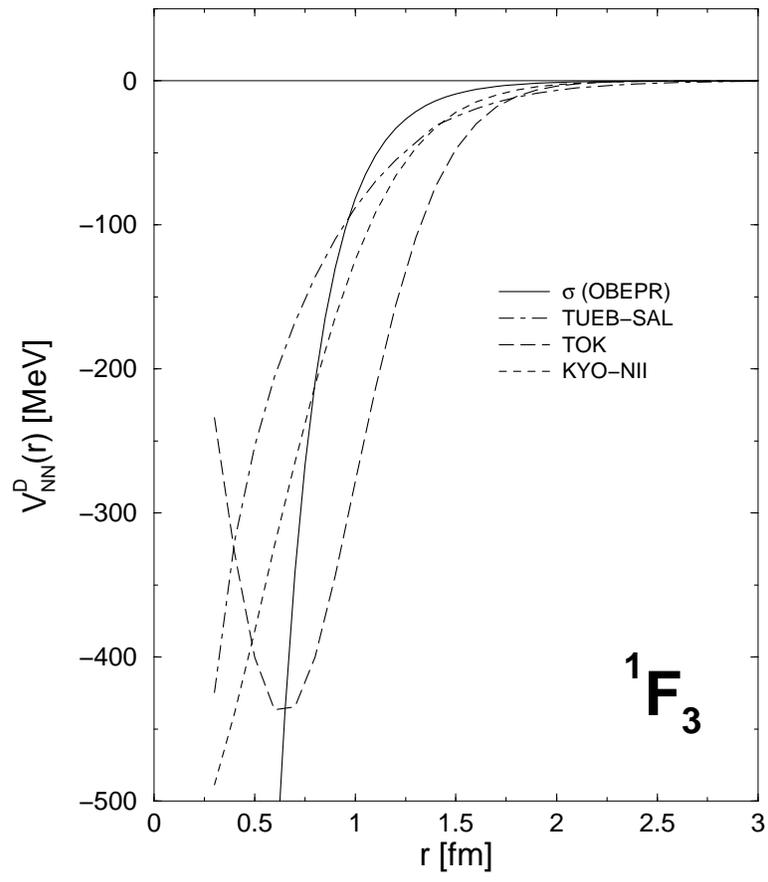
\begin{figure}
\caption{``Direct'' effective $NN$ interaction of the quark
models in the $^1F_3$ partial wave. Note that the spin-spin
part of the one-pion-exchange contribution is omitted. 
Same description of curves as in Fig.~\ref{phcut}.
The solid line shows the $\sigma$-exchange contribution of 
the Bonn OBEPR model.}
\label{po1f3} 
\end{figure}

In order to get a deeper understanding of the phase-shift
results let us examine the different quark-model potentials in 
coordinate space. Corresponding graphs are presented in Fig.~\ref{po1f3} 
for the $^1F_3$ partial wave. Note that the curves do not include
the contributions from the spin-spin part of the pion exchange.
These are practically the same in all considered $NN$ models
and therefore not interesting. Thus, Fig.~\ref{po1f3} displays 
the ``true'' medium-range part of the quark models. As discussed 
in the previous section, this part is generated by $\sigma$ exchange 
and/or by $\sigma$-like 
phenomenological terms. Accordingly, we expect that it should correspond 
roughly to the $\sigma$-exchange contribution that is present in 
conventional OBE models. However, a comparison with the 
$\sigma$ exchange of the Bonn OBEPR model
(cf. the solid line in Fig.~\ref{po1f3}) reveals that the latter
is significantly smaller than the corresponding pieces in 
the quark models - for internucleonic distances $r\geq$ 1 fm relevant 
for the $^1F_3$ partial wave. As a matter of fact, the medium-range 
part in the quark models is not only larger but seems to be longer 
ranged as well. 
In particular, the $\sigma$-like piece of the TOK potential 
(dashed curve) turns out to be exceptionally large. In view of this
it is not surprising that the corresponding phase shifts deviate
so strongly from the experimental results. On the other hand,
the KYO-NII model, which comes closest to the $\sigma$ exchange 
in OBE model gives also the best results for $^1F_3$ among the
quark models.

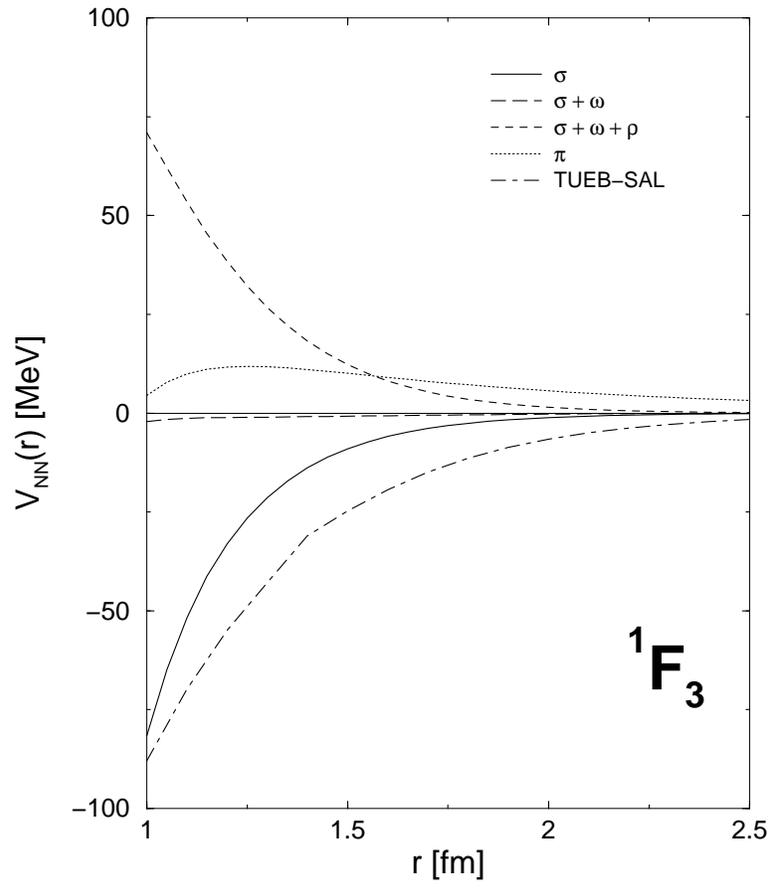
\begin{figure}
\caption{Contributions to the potential in the $^1F_3$ partial wave
for the one-boson-exchange model OBEPR. 
$\sigma$ exchange: solid line; 
$\sigma + \omega$ exchange: long-dashed line; 
$\sigma + \omega + \rho $ exchange: short-dashed line; 
$\pi$ exchange: dotted line. 
The dash-dotted curve shows the ``direct'' effective $NN$ interaction 
of the T\"ubingen-Salamanca model \protect\cite{TUEB1}, cf.
Fig.~\ref{po1f3}.} 
\label{po1f3b} 
\end{figure}

At this point let us recall that conventional meson-exchange
model such as OBEPR contain further ingredients that contribute to
the potential at medium-range distances, namely the exchanges of
the vector mesons $\rho$ and $\omega$. (Note that OBEPR contains
also contributions from $\eta$ and $a_0$ exchanges. However, their
effect on the higher partial waves that we discuss here is 
negligibly small and therefore we don't consider them explicitly.)
As mentioned already above, in the quark models of the $NN$ interaction
contributions from vector-meson exchange are left out altogether, 
as is argued, for conceptional reasons~\cite{Yaz}. Repulsive contributions, 
provided in the conventional meson-exchange models predominantly
by the $\omega$ exchange, are present in the quark models too. 
Here they are generated, in general, by OGE in conjunction
with quark exchange between the nucleons. However, this 
mechanism is rather short ranged and therefore does not contribute
to $F$ and higher partial waves anymore, as we have shown in the
last section. Consequently, for the quark models the $\sigma$-like 
contributions shown in Fig.~\ref{po1f3} constitute already the 
complete potential for medium-range distances. In conventional
meson-exchange models such as OBEPR the situation is different, as
can be seen in Fig.~\ref{po1f3b}. In this figure we show the potential
resulting from the $\sigma$ exchange (solid line) and then add
consecutively the contributions from $\omega$ and $\rho$ exchange.
The $\omega$-meson exchange practically cancels the attractive 
contribution from the $\sigma$ meson (long-dashed line).
Adding the $\rho$ meson (which is also repulsive in this partial wave) 
leads to a final result for the medium-range contributions 
which is repulsive (short-dashed line). The spin-spin part of the 
OPE - indicated by the dotted line - is repulsive as well. 
Combining those two leads to a strongly repulsive potential that
produces phase shifts as required by the data. 
In the case of the quark models the complete medium-range contributions 
are always attractive, cf. Fig.~\ref{po1f3} (the result for the TUEB-SAL 
model is also shown in Fig.~\ref{po1f3b} for the ease of comparison). 
Thus, they will reduce the repulsion provided by the pion-exchange
tail instead of enhancing it. In fact, for all models the attraction
increases rather strongly when going to shorter distances and,
consequently, eventually the whole potential becomes attractive. 
This feature is reflected in the behavior of the phase-shift results 
- which all turn to positive values for higher energies. 

We consider the above results as evidence that vector mesons still
play an important role in the $NN$ interaction at medium-range
distances. Present-day quark-model descriptions lack contributions 
of the range and strength as provided by the $\omega$ and $\rho$ mesons 
in OBE models. 

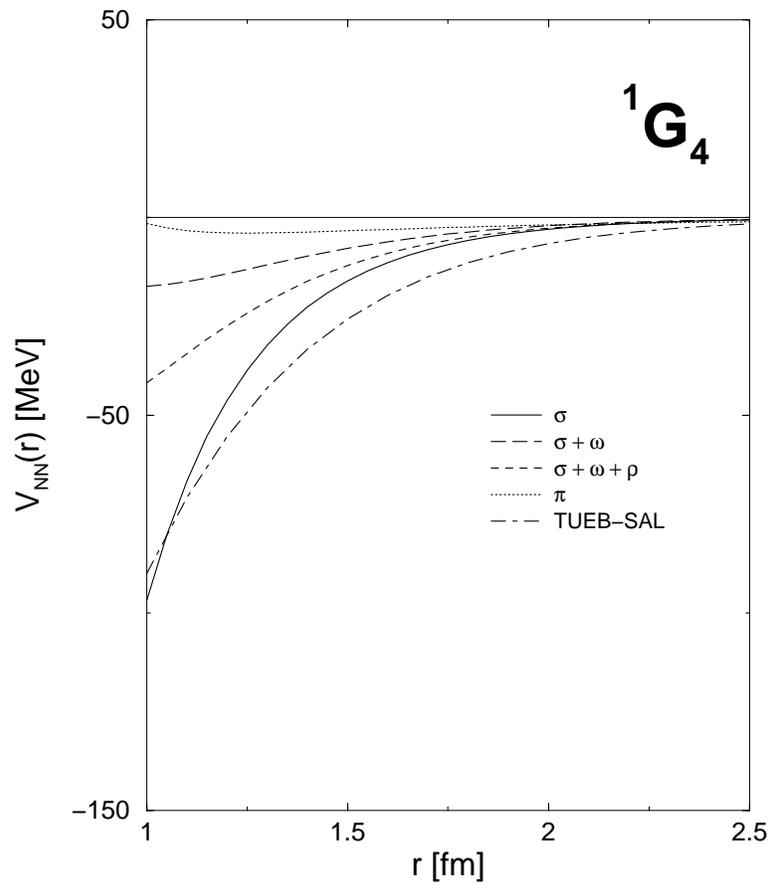
\begin{figure}
\caption{Same as in Fig.~\ref{po1f3b} for the $^1G_4$ partial wave.}
\label{po1g4} 
\end{figure}

Let us now look at the situation for the $^1G_4$ partial wave. 
Corresponding results are shown in Fig.~\ref{ph1g4}. Obviously,
besides OBEPR also the quark model KYO-NII is in good agreement
with the phase shift analysis. The other quark models either overshoot 
the experimental data (TUEB-SAL) or yield an underestimation (TOK).
Also here it is instructive to look at the various contributions to
the potential, which is done in Fig.~\ref{po1g4}. Again, we see that
the medium-range component of the quark model (TUEB-SAL; dash-dotted line)
is stronger and longer ranged than the $\sigma$ exchange contribution
in the OBE potential (solid line). Moreover, in the OBE model 
there is again a non-negligible contribution from the exchange of
vector mesons. However, since the $^1G_4$ partial wave is in a
different isospin channel, now the contributions from the iso-vector
mesons ($\rho$, $\pi$) have the opposite sign. As a consequence
the potential resulting from the $\omega$ exchange cancels to a
large extent with the one resulting from the $\rho$ exchange. 
Thus, the total medium-range contributions are pretty close to the
contributions of the $\sigma$ exchange alone (cf. the short-dashed
and solid lines). This fact that the contributions of the 
vector-meson exchange basically cancel out in this particular
partial wave is certainly responsible for the good performance
of some quark models, specifically of the model KYO-NII. 
In case of the TUEB-SAL model the $\sigma$-exchange contribution
is simply too strong and long-ranged and therefore
the phase shifts are too large. For the TOK model the situation
is somewhat different. The $\sigma$-like component of
this potential has a phenomenological part whose parameters are 
adjusted for each of the four spin-isospin ($S,T$=0,1)
channels separately, cf. sect. 2.3. of Ref.~\cite{TOK} for details.
For the $^1F_3$ partial wave [(0,0) channel] this phenomenological
piece is rather strong as we have seen above whereas for $^1G_4$
[(0,1) channel] it is much weaker. In addition, the TOK model 
contains a quadratic spin-orbit term 
of the form $-\bar V_{QSO}\,\bsigma\bcdot\bsigma\,\bL^2$, 
which provides strong repulsion in singlet states with high
orbital angular momentum $L$ such as the $^1G_4$.

\section{Discussion and Conclusion}

In the last section we have seen that many of the presently
available quark models of the $NN$ interaction have serious 
deficiencies in the description
of higher partial waves. Specifically, we have shown that those
models provide, in general, much too attractive forces at larger 
internuclear distances. A first possible and plausible explanation for 
these deficiencies was presented by K. Holinde already several years 
ago~\cite{Karl}. He argued that the defect of those quark models 
might result from the fact that the entire repulsive contributions 
are generated by gluon exchange alone and, therefore, are of 
extremely short-ranged nature. As a remedy he advocated that 
at least part of the long-range tail of the $\omega$-exchange 
from the standard meson-exchange picture should be kept in those
quark models. 

Our detailed investigations revealed that the above conjecture is
only {\it one} part of the truth. We confirmed that the repulsion provided
by the quark models is much too short-ranged and therefore does not
affect the higher partial waves anymore as it would be required for a
proper description of the corresponding phase shifts. However, the
situation is more complex. We found evidence that, besides the
$\omega$ exchange, also the long-range tail of the $\rho$ meson
exchange is still felt by the $F$ and $G$ waves and therefore needed
for a quantitative reproduction of those phases. As already pointed
out above, contributions from those vector mesons are left out in
the quark models from the very beginning - and there are no mechanisms
in those models that would generate forces with similar features and
comparable range. Finally, and most disturbingly, we found that most
of the quark models contain attractive ($\sigma$ like) contributions
that are rather strong and also rather long-ranged. 

The reason why such strong attractive forces need to be introduced
in the quark models would require a thorough analysis of the 
short-range part of those models which is beyond the scope of the
present study. Therefore, here we restrict ourselves to a plausible 
speculation that certainly deserves further detailed study. We believe 
that the origin of this defect are the difficulties which these quark 
models have in providing a sufficiently strong spin-orbit force for 
describing the splitting of the spin-triplet $P$ waves ($^3P_0$, $^3P_1$, 
$^3P_2$)~\cite{TUEB1}. These spin-orbit forces are either generated by 
one-gluon exchange and/or by the $\sigma$ exchange between quarks. 
Since the spin-orbit force provided by the one-gluon exchange is
very weak as compared to the central piece one has to introduce a
large coupling constant in order to achieve sufficient spin-orbit
force and that, in turn, leads to a huge repulsive central contribution.
Agreement with the experimental phase shifts can then only be achieved 
by introducing a like-wise huge attractive central ($\sigma$-like) 
piece that counter-balances this strong repulsion. Those two 
ingredients can be adjusted in such a way that they compensate very
well for the lower partial waves. But this does not work anymore
for the higher partial waves because of the different ranges 
involved in these contributions. On the other hand, 
if the spin-orbit force is generated by the $\sigma$ exchange alone
this contribution has to be made stronger than in conventional meson 
exchange models as well, because in the latter model one also gets 
additional and significant contribution to the spin-orbit force 
from $\omega$ exchange. As pointed out already above, such contributions
are left out in the quark models. 

For obvious reasons the free parameters in those quark models have
been adjusted to give a good description of the lower (i. e. $S$, 
$P$, and $D$) partial waves. But this procedure automatically 
fixes the medium-range (or meson-exchange) part of the $NN$ force 
and, consequently, the predictions of those models for the higher 
partial waves. Our investigations have shown that the meson-exchange 
part of the quark models is 
not realistic yet but rather in conflict with the present-day knowledge 
about the medium- and long-range properties of the $NN$ force 
obtained from other sources. Thus, we confirm a conjecture that was
already raised in Ref.~\cite{TOK}. At the same time we want
emphasize, however, that one should be careful with the second part 
of the conjecture stated in Ref.~\cite{TOK}, 
namely that the failure in describing the higher partial waves is 
not caused by a problem in the short-range part (i. e. the part of 
the $NN$ interaction that depends on the quark degrees of freedom),
for the following reason:
Low partial waves like $S$ waves feel the short-range part  
of the $NN$ interaction as well as the medium- and long-range parts.  
Thus, if the short-range part of the $NN$ force derived in those 
quark models still has deficiencies, it might be possible to conceal 
those at the expense of introducing large and unrealistic
medium-range components into the $NN$ model in a more-or-less
phenomenological way. Of course, then these deficiencies could show 
up indirectly and somewhere else, namely in unrealistic predictions 
for the higher partial waves. 

Therefore, we suggest that one should follow a different strategy
if one wants to test the short-range part of the $NN$ force as
derived from subnucleonic degrees of freedom. One should
include our knowledge on the medium- and long-range parts of the
$NN$ interaction from the very beginning and use it as a constraint
for the $NN$ model to be constructed. Reliable results for the
$NN$ interaction at intermediate ranges have been derived in 
the past, for example, from dispersion theory \cite{Paris-NN} as well 
as in an extended meson-exchange model \cite{Bonn} and more
recently in the context of chiral perturbation 
theory \cite{NNchR,NNchK}. These pieces of information should be utilized
and supplemented with the short-range piece of the $NN$ interaction
as it emerges from the quark-model picture. We believe that only
by following this procedure solid and conclusive results about
the quality and reliability of a quark-model description of the 
short-range part of the $NN$ interaction can be achieved. 

\acknowledgments{The authors thank Dr. Y. Fujiwara for sending us the phase 
shifts of the KYO-NII potential. Private discussions with A. Faessler, 
F. Fern\'andez, A. Buchmann and A. Valcarce are acknowledged. Financial 
support for this work was provided in part by the international exchange 
program DLR (Germany, BRA W0B 2F) - CNPq (Brazil, 910133/94-8).}

\newpage 

\newpage

\vglue 1cm 
\begin{figure}
\centerline{\epsfxsize=10.0cm\epsfbox{ph_com.epsi}}
\end{figure}

\vskip 4cm

\center{Fig. 1}

\newpage
\vglue 1cm

\begin{figure}
\centerline{\epsfxsize=10.0cm\epsfbox{cut1f.epsi}}  
\end{figure}

\vskip 4cm

\center{Fig. 2}

\newpage
\vglue 1cm

\begin{figure}
\centerline{\epsfxsize=10.0cm\epsfbox{ph1f3.epsi}}
\end{figure}
 
\vskip 4cm

\center{Fig. 3}

\newpage
\vglue 1cm

\begin{figure}
\centerline{\epsfxsize=10.0cm\epsfbox{ph1g4.epsi}}
\end{figure}

\vskip 4cm

\center{Fig. 4}

\newpage
\vglue 1cm

\begin{figure}
\centerline{\epsfxsize=10.0cm\epsfbox{poq1f.epsi}}
\end{figure}
 
\vskip 4cm

\center{Fig. 5}

\newpage
\vglue 1cm

\begin{figure}
\centerline{\epsfxsize=10.0cm\epsfbox{pod1f.epsi}}
\end{figure}
 
\vskip 4cm

\center{Fig. 6}

\newpage
\vglue 1cm

\begin{figure}
\centerline{\epsfxsize=10.0cm\epsfbox{pod1g.epsi}}
\end{figure}

\vskip 4cm

\center{Fig. 7}

\begin{thebibliography}{99}
\bibitem{Bonn} 
R. Machleidt, K. Holinde and Ch. Elster, Phys. Rep. {\bf 149}, 1 (1987).
\bibitem{Bonn2} 
R. Machleidt, Adv. Nucl. Phys. {\bf 19}, 1 (1988). 
\bibitem{LeYaouanc} For a list of references, see: A. Le Yaouanc, Ll. Oliver, 
O. P\`ene and J.-C. Raynal, Hadron Transitions in the Quark Model (Gordon and 
Breach, New York, 1988).
\bibitem{Libermann} D.A. Liberman, Phys. Rev. D {\bf 16}, 1542 (1977).
\bibitem{Neudatchin} V.G. Neudatchin, Y.F. Smirnov, and R. Tamagaki, Prog. 
Theor. Phys. {\bf 58}, 1072 (1977); V.G. Neudatchin, Y.F. Smirnov, and
Y.M. Tchuvil'sky, Phys. Lett. {\b 88B}, 231 (1979).  
\bibitem{Harvey} M. Harvey, Nucl. Phys. {\bf 352}, 326 (1981).
\bibitem{Ribeiro} J.E.F.T. Ribeiro, Z. Phys. C {\bf 5}, 27 (1980).
\bibitem{WarkeShanker} C.S. Warke and R. Shanker, Phys. Rev. C {\bf 21}, 2643 
(1980).
\bibitem{OkaYazaki} M. Oka and K. Yazaki, Phys. Lett. {90B}, 41 (1980);
Prog. Theor. Phys. {\bf 66}, 551 (1981); 
{\bf 66}, 5572 (1981).
\bibitem{Faessler} A. Faessler, F. Fernandez, G. L\"ubeck and K. Shimizu, 
Phys. Lett. {\bf 112B}, 201 (1982); Nucl. Phys. {\bf A402}, 555 (1983).   
\bibitem{TUEB-SAL} F. Fern\'andez, A. Valcarce, U. Straub, and A. Faessler,
J. Phys. G {\bf 19}, 2013 (1993); A. Valcarce, A. Buchmann, F. Fern\'andez, and
A. Faessler, Phys. Rev. C {\bf 50}, 2246 (1994). 
\bibitem{TUEB1} A. Valcarce, A. Buchmann, F. Fern\'andez, and
A. Faessler, Phys. Rev. C {\bf 51}, 1480 (1995). 
\bibitem{SAL-VAL}  A. Valcarce, P. Gonz\'alez, F. Fern\'andez, and V.
Vento, Phys. Lett. {\bf B367}, 35 (1996); A. Valcarce, F. Fern\'andez, and 
P. Gonz\'alez, Phys. Rev. C {\bf 56}, 3026 (1997); Few-Body Systems Suppl. 
{\bf 99}, 1 (1998).
\bibitem{TOK} S. Takeuchi, K. Shimizu, and K. Yazaki, 
Nucl. Phys. {\bf A504}, 777 (1989).
\bibitem{KYO-NII} Y. Fujiwara, C. Nakamoto, and Y. Suzuki, Phys. Rev. Lett.
{\bf 76}, 2242 (1996); Phys. Rev. C {\bf 54}, 2180 (1996).
\bibitem{Yaz} K. Yazaki, Prog. Part. Nucl. Phys, {\bf 24}, 353 (1990).
\bibitem{Paris-QM} R. Vinh Mau, C. Semay, B. Loiseau and M. Lacombe, Phys. Rev.
Lett. {\bf 67}, 1392 (1991).
\bibitem{Paris-NN} M. Lacombe, B. Loiseau, J.M. Richard, R. Vinh Mau, 
J. Cot\'e, P. Pires and M. de Tourieil, Phys. Rev. C {\bf 21}, 861 (1980).   
\bibitem{nijm} V.G.J. Stoks, R.A.M. Klomp, C.P.F. Terheggen, and
J.J. de Swart, Phys. Rev. C {\bf 49}, 2950 (1994). 
\bibitem{cdbonn} R. Machleidt, {\tt nucl-th/0006014} (2000).
\bibitem{exper1} V.G.J. Stoks, R.A.M. Klomp, M.C.M. Rentmeester, and
J.J. de Swart, Phys. Rev. C {\bf 48}, 792 (1993). 
\bibitem{exper2} R.A. Arndt, J.S. Hyslop III, and L.D. Roper,  
Phys. Rev. D {\bf 35}, 128 (1987). 
\bibitem{exper3} D.V. Bugg and R.A. Bryan, Nucl. Phys. {\bf A540}, 
449 (1992). 
\bibitem{Karl} K. Holinde, Nucl. Phys. {\bf A543}, 143c (1992).
\bibitem{NNchR} M.R. Robilotta and C.A. da Rocha, Nucl. Phys. {\bf A615},
391 (1997); J.-L. Ballot, M.R. Robilotta, and C.A. da Rocha,
Phys. Rev. C {\bf 57}, 1574 (1998).
\bibitem{NNchK} N. Kaiser, R. Brockmann, and W. Weise, Nucl. Phys. 
{\bf A625}, 758 (1997); N. Kaiser, S. Gerstend\"orfer, and W. Weise, Nucl. 
Phys. {\bf A637}, 395 (1998).  
\end{thebibliography}
\end{document}